\documentclass[twocolumn,showpacs,preprintnumbers,amsmath,amssymb]{revtex4}
\usepackage{amssymb} \usepackage{graphicx}
\usepackage{dcolumn} \usepackage{amsmath}
\usepackage[latin5]{inputenc}

\begin{document}
\title{Reduced quantum dynamics with
initial system-environment correlations characterized by pure Markov states}

\author{A. Türkmen $^{1,2}$}
 \email{aturkmen@ankara.edu.tr}
\author{A. Verçin $^{1}$}
 \email{vercin@science.ankara.edu.tr}
\author{S. Yılmaz$^{1,3}$}
\email{soyilmaz@ankara.edu.tr}
%\thanks{E.mail:vercin@science.ankara.edu.tr}%
\address{$^{1}$ Department of Physics, Ankara University, Faculty of Sciences,
06100, Tandoğan-Ankara, Turkey\\
$^{2}$ Department of Physics, Ordu University, Faculty of Sciences and Arts,
52200, Ordu, Turkey\\
$^{3}$Department of Physics, Amasya University, Faculty of Sciences and Arts, 05100, Amasya,
Turkey.}
\date{\today}
\begin{abstract}
Any tripartite state  which saturates the strong subadditivity relation for the
quantum entropy is defined as the Markov state. A tripartite pure state describing
an open system, its environment and their purifying system is a
pure Markov state
iff the bipartite marginal state of the purifying system and environment is a product
state. It has been shown that as long as the purification of the input
system-environment state is a pure Markov state
the reduced dynamics of the
open system can be described, on the support of initial system state, by a
quantum channel for every joint unitary evolution of the system-environment
composite even in the presence of initial correlations. Entanglement, discord
and classical correlations of the initial system-environment states implied by
the pure Markov states are analyzed and it has been shown that all
these correlations are entirely specified by the entropy of environment. Some
implications  concerning perfect quantum error correction procedure and quantum
Markovian dynamics are presented.
 \end{abstract}
 \maketitle

\section{Introduction}
Each real world  quantum system forms a
closed compound system with its
surrounding environment and evolves together with it under
 joint unitary evolutions. Such
systems are called  open quantum systems (OQSs) \cite{Breuer-Petruccione2002,Schlosshauer2008}. Each OQS interacts and
gets correlated to some extent with its environment and evolves according to quantum rules
individually, at least for
a while, before completely loosing its quantum coherence
property \cite{Schlosshauer2008}. On the other hand, the main goal in
realizing many quantum information and quantum computation tasks is to
maintain, as long as possible, the coherence properties of information carriers which are
constantly interacting with ambient medium \cite{NC}. However, despite their importance in our understanding the
quantum aspects of the nature around us and in the emerging fields of quantum technologies we
still lack a complete understanding of evolutions of OQSs initially correlated
with their environment.

Physically most appealing way to describe evolutions of
an OQS is by the so-called completely positive (CP) maps \cite{Stinespring,Kraus}.
These are linear
maps which transform
every positive operators in their definition domain
to positive operators and maintain this
property in all tensorial
extensions. Trace preserving
CP maps are called quantum channels and they are interchangeably
referred to also as CPTP maps: These map any quantum state (density operator; positive operator with
unit trace) to another quantum state \cite{Holevo}. To emphasize one of
the physical
intuitions behind such maps  let us consider a given CPTP map acting on an OQS.
By appending
an auxiliary system, say the environment of OQS,
the tensorial extension of such a map can be defined on any joint quantum
state  of the compound system, the OQS and auxiliary system. The
result of extended action is certainly another admissible joint state
and this is what complete positivity corresponds to in applications. Moreover,
 when the effects of
appended system
are then averaged out what remains is the action of the same CPTP map on
the reduced state
of the original OQS. This holds irrespective of the correlations
the initial joint state may have
and of the dimension of added system provided that
the original state of OQS is the reduced state of the joint state.

In real world
and in the laboratory applications however CP maps
are reduced
from the joint unitary evolutions and the essential problems arise
in this context. In such a case the action of a joint unitary map on a joint state
may not give the action of a even positive map, let alone CP map, on
the reduced state of the OQS after
discarding the environment \cite{Pechukas94}
(see also \cite{Sudarshan 2008}). Certainly,
 uncorrelated joint states, that is, product states are
 exceptions and starting with such a
state has become a basic assumption in
almost all approaches to the
dynamics of OQSs.

In fact,  there is a whole set of exceptions that provides a large
family of initially correlated joint states,
not recognized in the literature
before the recent work \cite{Buscemi2014}. The main goal of
our study is to specify such a well defined special subset of correlated
initial system-environment states that not only permits CP reduced
dynamics for the observed OQS, but also makes it possible  to
characterise  all classical and quantum correlations of its elements.
The states that will be explored here are tripartite states that can
be reconstructed from their marginal states via the actions of CP maps.
Since averaging the effect of a subsystem is carried out by a partial
trace, such a CP map
locally reverses the action of  partial trace map on the considered
state. Evidently,
a product
 joint state is such a state since tensoring by an
additional state is a CPTP map on the other factor and therefore
this family  contains uncorrelated initial states as special cases.

Reconstruction positive maps were also known in the context of OQSs
 under the name of
the assignment maps but, unfortunately, they were not explored
 sufficiently enough. A detailed study
 of reconstruction CP maps were made in another context; in characterising
 tripartite states
 that saturate
 the strong subadditivity (SSA) relation for the (von Neumann)
 quantum entropy \cite{Petz2003,Petz2004,Ruskai2002} (and the references therein). In this
 context such a map is known as
the Petz recovery  map and related tripartite
 states are called Markov states. Very recently, by
including an additional blind and dead reference system into discussion F. Buscemi
has shown in Ref. \cite{Buscemi2014} that the reduced dynamics of
 an OQS can be described by CP maps
in the presence of initial correlations.
Such a description is possible for
tripartite states of the reference system, the OQS and the
environment trio for which the quantum mutual information
between the reference system and the
environment, conditional on the system, is zero. (Positivity of
this mutual information is better known as the SSA relation \cite{Petz2004,NC}.)
It should be noted that the system-environment states which are marginals
of tripartite input Markov states, conditional on system are not the only
 states for which the system evolution is given by a
CPTP linear map.

In this study, we shall use a
similar tripartite framework of Ref. {\cite{Buscemi2014}} but
 for a detailed exposition
of the problems and in order to be able
to analyze the initial correlations explicitly, we shall restrict our consideration
mainly to
pure Markov states.
A direct and detailed study of these states and of related quantum channels
as well as qualitative and
quantitative characterizations
of all possible classical and quantum correlations of the initial
system-environment
states are, to the best of our knowledge, new contributions of this
study to the present OQS literature.
\section{Framework of this study and summary}
We shall denote the OQS and
its environment, respectively, by
$Q$ and $E$ and suppose that the bipartite system
$QE$ forms a closed system subject to time-dependent joint unitary
operator $U^{QE}$. The purifying system of $QE$ states will be
represented by $R$. Each system is supposed to be endowed with a
finite dimensional Hilbert space ${\cal H}_X$ and with
the space $B({\cal H}_X); X=R, Q, E$, of bounded operators.
If we denote the dimension of ${\cal H}_X$ by $d_X$, then $d_R$
is not smaller than  $d_Qd_E$.

Even
when $Y$ represents
a compound system, its quantum states will be denoted by density operators
$\rho^Y$. The tensorial extension to
$B({\cal H}_Q\otimes {\cal H}_E)=B({\cal H}_{QE})$ of a map $\Lambda$
defined on $B({\cal
H}_Q)$ will be denoted by $\Lambda \otimes id_{E}, id_X$ being the
identity map
 on $B({\cal H}_X)$. If $\Lambda$ is a
 positive map and if $\Lambda \otimes id_{E}$ preserves the positivity of operators
 defined on $B({\cal H}_{QE})$ for all dimensions of ${\cal H}_E$, $\Lambda$ is a CP map.
Partial traces and adjoint actions
of unitary operators are the standard examples of CPTP maps.

Throughout this study the initial and final states will be denoted by indexed
$|\psi\rangle, \rho$
and $|\phi\rangle,\sigma$, respectively. Superscripts over them will indicate which system they belong to.
They should be thought of also as indexed
by an initial time $\tau$ and final time $\tau^{\prime}\geq \tau$.
Accordingly,
$U^{QE}$ and  related channels should be thought of as indexed by both times leading to
state vectors and operators at the initial time to that of at
the final time. At the beginning $QE$ is supposed to be in the reduced state
$\rho^{QE}= Tr_R|\psi^{RQE}\rangle\langle
\psi^{RQE}|$
of the initial tripartite pure state
$|\psi^{RQE}\rangle$. Evidently the rank of $\rho^{R}$, that is,
the number of nonzero eigenvalues of
$\rho^{R}=Tr_{QE}|\psi^{RQE}\rangle\langle
\psi^{RQE}|$,  is always equals to the rank of $\rho^{QE}$: $rank(\rho^{QE})=rank({\rho^{R}})$.

Defining
the adjoint action of an operator $V$ on
$\rho$ by $ad_V(\rho)=V\rho V^\dag$, the output state is
$\sigma^{RQE}
=(id_R\otimes ad_{U^{QE}})\rho^{RQE}$. Hence $R$ remains blind and dead during the evolution.
Since the overall evolution is unitary, when the input state
is pure $\rho^{RQE}=|\psi^{RQE}\rangle\langle \psi^{RQE}|$ then so is the
output $\sigma^{RQE}= |\phi^{RQE}\rangle\langle
\phi^{RQE}|$ where $|\phi^{RQE}\rangle=(\mathbb{I}_R\otimes
U^{QE})|\psi^{RQE}\rangle$ ($\mathbb{I}_X$ stands for the unit operator, or
the unit matrix of ${\cal H}_X$). In any case, the
reduced dynamics of $Q$ is specified by tracing out the environment and
the purifying system:
\begin{eqnarray}
\sigma^{Q}={\cal E}(\rho^{Q})
=Tr_{RE} (\sigma^{RQE}).
\end{eqnarray}
Now the important question is that, for what kind of
initial correlations of $\rho^{QE}$ is the map
${\cal E}$ a linear CPTP map?

As a non-exhaustive answer to the above question in what follows we
shall prove that
as long as the input $QE$ state is a reduced state of
a pure Markov state
the evolution map ${\cal E}$ is a
CPTP map for every joint unitary
evolution of the $QE$ composite in the presence
of initial correlations
implied by the pure Markov state. This is shown in Sec. IV where
the explicit form of the channel and its Kraus operators as well as
identification some special cases of channel
are presented. In Sec. III the necessary entropy relations, Markov states, Petz map and
pure Markov states are introduced. Canonical form of the pure Markov states
and their characteristic
traits are
also exhibited in Sec. III. The correlations such as entanglement
of formation, discord and
classical correlations that the initial $QE$ states may have are analyzed in
Sec. V.
There it is shown that the
entropy of the environment
entirely specifies all these correlations. Our main points concerning the CPTP evolutions and
characterizations of initial
correlations are summarized by two theorems. In the final section
intimate connections of our results with
perfect quantum error correction procedure
and quantum Markovian dynamics are discussed.

\section{SSA relation, Markov states, Petz map and Pure Markov states}
The von Neumann entropy
of a state $\rho^Y$ is defined by $S(\rho^Y)=-Tr\rho^Y\log \rho^Y$: This will be
denoted simply by $S(Y)=S(\rho^Y)$. The quantum conditional
entropy $S(X|Y)=S(XY)-S(Y)$ and the quantum
mutual information $S(X;Y)$ defined by
\begin{equation}
S(X;Y)=S(X)+S(Y)-S(XY),
\end{equation}
will be distinguished with special punctuation inside the parenthesis. $S(X;Y)$ is zero iff
$\rho^{XY}$ is the product state $\rho^{XY}=\rho^X\otimes \rho^Y$, where
$\rho^X=Tr_Y\rho^{XY}$ and $\rho^Y=Tr_X\rho^{XY}$ are the marginal states
of $\rho^{XY}$.

Accordingly, the conditional mutual
information $S(R;E|Q)$, conditioned on $Q$, for a state of
a tripartite system $RQE$ is defined as $S(R;E|Q)=S(R|Q)+S(E|Q)-S(RE|Q)$. In view of the
definition of conditional entropy this takes the form
\begin{eqnarray}
 S(R;E|Q)=S(RQ)+S(QE)-S(RQE)-S(Q),
\end{eqnarray}
and the celebrated SSA relation which hosts several entropy relations can be expressed by
$S(R;E|Q)\geq 0$.

A tripartite state is a Markov state conditional on $Q$ iff it satisfies
$S(R;E|Q)=0$. A key property of Markov states we shall use is that: $\rho^{RQE}$ is a Markov
state iff there exists a CPTP map ${\cal R}: Q\rightarrow QE$ such that
\begin{eqnarray}
\rho^{RQE}=(id_R\otimes {\cal R})\rho^{RQ}
,\nonumber
\end{eqnarray}
where $\rho^{RQ}=Tr_E\rho^{RQE}$
(Eq. (11) of Ref. \cite{Petz2004} see also \cite{Petz2003}). On
the support of $\rho^{Q}=Tr_{RE}\rho^{RQE}$ the action of
${\cal R}$ on any $X\in B({\cal H}_Q)$ is given by (Eq. (15) of Ref. \cite{Fawzi-Renner2014})
\begin{eqnarray}
 {\cal R}(X)=ad_{(\rho^{QE})^{1/2}}\big[\big(ad_{(\rho^{Q})^{-1/2}}X\big)
 \otimes \mathbb{I}_E\big].
\end{eqnarray}
 Note that for $X=\rho^{Q}$ we have
${\cal R}(\rho^{Q})=\rho^{QE}$. That is, ${\cal R}$ locally reverses
the action of  $Tr_E$ on $\rho^{QE}:\;({\cal R}\circ Tr_E)(\rho^{QE})=\rho^{QE}$, where
$\circ$ denotes the composition of maps.

Henceforth ${\cal R}$ will be referred to as the Petz map. As is apparent
from  Eq. (4), ${\cal R}$ can be considered to be a composition of two $CP$ maps: The first is of the form
$B({\cal H}_Q)\rightarrow B({\cal H}_{QE})$ and is defined by
$X \rightarrow (ad_{(\rho^{Q})^{-1/2}}X\big)
 \otimes \mathbb{I}_E$ and the second is  of the form
 $B({\cal H}_{QE})\rightarrow B({\cal H}_{QE})$
and is defined by $Y\rightarrow ad_{(\rho^{QE})^{1/2}}(Y)$ for any
$Y\in B({\cal H}_{QE})$. By virtue of
 the general relation $Tr_E[Y(X\otimes\mathbb{I}_E )]=(Tr_EY)X$ one can
 easily verify that
 \begin{eqnarray}
 Tr_{QE}[{\cal R}(X)]&=&
 Tr_Q\Big\{(Tr_E\rho^{QE})[(\rho^{Q})^{-1/2}X(\rho^{Q})^{-1/2}
 ]\Big\}\nonumber\\
 &=&
 Tr_Q X.\nonumber
\end{eqnarray}
Thus, the Petz map  ${\cal R}$ is indeed a quantum channel on the support of $\rho^{Q}$.
As it depends on the initial $\rho^Q$ state, ${\cal R}$ should be indexed by
$\rho^Q$, but for the sake of clarity this dependence is suppressed.
When extended to all of $B({\cal H}_Q)$ the Petz map can be considered as a
trace-non-increasing CP map \cite{Wilde2015}.

\subsection{Pure Markov states}
For any tripartite pure state
$S(RQE)$ vanishes and according to Schmidt
decomposition the bipartite splits $R|QE, RQ|E$ and $Q|RE$ imply the following equalities;
\begin{eqnarray}
 S(R)=S(QE),\quad S(E)=S(RQ),\quad S(Q)=S(RE).\nonumber
\end{eqnarray}
Substituting these relations into Eq. (3) immediately
proves the following statement.

\textit{Lemma 1.} For any tripartite pure state the equality
$S(R;E|Q)=S(R;E)$ holds. $\square$

That is, the conditional mutual information of any tripartite pure state, given $Q$, is just
the mutual information of $R$ and $E$. Since being a product state is
the necessary and sufficient conditions
for quantum mutual information of a
given bipartite state to vanish, as a
corollary of \textit{Lemma 1} we have the following fact which is a general trait of all pure
Markov states.

\textit{Corollary 1.} Any tripartite pure state $\rho^{RQE}=|\psi^{RQE}\rangle\langle
\psi^{RQE}|$ is a Markov state iff the marginal state $\rho^{RE}=Tr_Q\rho^{RQE}$ is the
product state $\rho^{RE}=\rho^R\otimes \rho^E$. $\square$

\subsection{Canonical form of the pure Markov states}
By
\textit{Corollary 1}, all tripartite pure Markov states of RQE are
purifications of product states of R and E.
To say more, let us consider the product state $\rho^{RE}=\rho^R\otimes
\rho^E$. Denoting the spectra of $\rho^R$
and $\rho^E$ by $\{\kappa_j\}$ and $\{\mu_k\}$
\begin{eqnarray}
\rho^{R}=\sum_{j}\kappa_j|r_{j}\rangle\langle r_{j}|,\quad \rho^{E}
=\sum_{k}\mu_k|\varepsilon_k\rangle\langle \varepsilon_{k}|,
\end{eqnarray}
with
the corresponding orthonormal
eigenstates $\{|r_j\rangle\}$ and $\{|\varepsilon_k\rangle\}$ we have \cite{Note1}
\begin{eqnarray}
 |\psi^{RQE}\rangle=\sum_{j,k}\sqrt{\kappa_j\mu_k}|r_j\rangle\otimes
 |q_{jk}\rangle\otimes|\varepsilon_k\rangle,
\end{eqnarray}
for purification of $\rho^R\otimes
\rho^E$. Here
$\{|q_{jk}\rangle ; \langle q_{mn}|q_{jk}\rangle=\delta_{jm}\delta_{kn}\}$
represents the set of
orthonormal eigenstates of the corresponding state of $Q$:
\begin{eqnarray}
\rho^{Q}=\sum_{j,k}\kappa_j\mu_k |q_{jk}\rangle\langle q_{jk}|.
\end{eqnarray}

In fact, any tripartite pure state $|\Phi\rangle$ whose
marginal $\rho^{RE}=Tr_Q|\Phi\rangle\langle \Phi |$ is a product state is, by definition, a
purification of $\rho^{RE}$. Moreover
 $|\Phi\rangle$ is unique up to local unitary (or, more generally,
  local isomorphism $V:{\cal H}^Q \rightarrow {\cal H}^{Q^\prime},
  V^\dag V=\mathbb{I}_Q$)
transformations of $Q$, that is
\begin{eqnarray}
\rho^{RE}=Tr_Q|\Phi\rangle\langle \Phi |= Tr_Q\big[(id_R\otimes ad_{V}\otimes
id_E)|\Phi\rangle\langle \Phi|\big].\nonumber
\end{eqnarray}
Thus Eq. (6) is a canonical form characterizing all pure Markov states and
from the explicit form of diagonal marginal states, or directly from (6)
we have the following statement.

\textit{Lemma 2.} For a given pure Markov state the ranks of its one-partite marginal
states satisfies  the equality:
\begin{eqnarray}
rank(\rho^{Q})=rank(\rho^{E})rank(\rho^{R}). \qquad \square
\end{eqnarray}

Some of the immediate corollaries of this Lemma can
be directly stated as follows. When both $R$
and $E$ are in pure states then so is $Q$ and
we have a pure Markov state as a pure product state.
When only one of $R$ and $E$ is in a pure state then the pure Markov state has, irrespective
of $rank(\rho^{Q})$, one of the following form;
\begin{equation}
\rho^{RQE}=\left\{
\begin{array}{c}
|\varphi^{RQ}\rangle\langle \varphi^{RQ}| \otimes |\psi^{E}\rangle\langle \psi^{E}|
,\\
\\
|\varphi^R\rangle\langle \varphi^R|
\otimes  |\psi^{QE}\rangle\langle \psi^{QE}|.
\end{array}
\right .
\end{equation}
When the $rank(\rho^{Q})$ is a prime number
then the above two forms exhaust
all possible forms of pure Markov states. In particular, Eqs. (9) exhibit all
possible pure Markov states
for  qubit and qutrit states of $Q$. Moreover,
for any tripartite pure state of $RQE$,
$\rho^{QE}$ is a pure state iff so is $\rho^{R}$. When $\rho^{R}$ is pure the tripartite pure
state is automatically a Markov state.  Hence the second relation of (9) exhausts the set of
pure Markov states in which $R$ is in a pure state.

In the most general case neither $QE$ nor $RQ$ needs to be in a pure state.
Indeed, in terms of orthonormal states
\begin{eqnarray}
|\psi_j^{QE}\rangle&=&\sum_{k}\sqrt{\mu_k}
  |q_{jk}\rangle\otimes|\varepsilon_k\rangle,\;
  \langle \psi_m^{QE}|\psi_j^{QE}\rangle=\delta_{mj},\nonumber\\
  \\
|\psi_k^{RQ}\rangle&=&\sum_{j}\sqrt{\kappa_j}
 |r_j\rangle\otimes|q_{jk}\rangle,\;
  \langle \psi_n^{RQ}|\psi_k^{RQ}\rangle=\delta_{nk},\nonumber
\end{eqnarray}
from Eq. (6) we obtain
\begin{eqnarray}
\rho^{QE}=\sum_{j}\kappa_j|\psi_j^{QE}\rangle\langle \psi_j^{QE}|,\quad
 \rho^{RQ}=\sum_{k}\mu_k|\psi_k^{RQ}\rangle\langle \psi_k^{RQ}|.\nonumber
\end{eqnarray}
Thus when neither $R$ nor $E$ is in pure state, both $\rho^{QE}$ and
$\rho^{RQ}$ are mixed.

\section{Pure Markov states and CP evolutions}
From now on we consider our pure tripartite
input state $\rho^{RQE}$ to be a pure Markov state obeying
\textit{Corollary 1};
$\rho^{RE}=\rho^R\otimes \rho^E$. To determine the output state we first
compute $\rho^{RQ}$ and write $\rho^{RQE}
=(id_R\otimes {\cal R})\rho^{RQ}$.
Then the tripartite output can be written as
\begin{eqnarray}
 \sigma^{RQE}&=&(id_R\otimes ad_{U^{QE}})
 \rho^{RQE}\nonumber\\
 &=&\big[id_R\otimes (ad_{U^{QE}}\circ {\cal R})\big]\rho^{RQ}.
\end{eqnarray}
By writing $\rho^{RQ}$ in the block form
$\rho^{RQ}=\Sigma_{ij}e_{ij}\otimes Q_{ij}$, where
$e_{ij}=|e_i\rangle\langle e_j|$
are the standard matrix units corresponding to the standard
unit vectors $|e_i\rangle$ of
${\cal H}_R$ and $\rho^{Q}=\Sigma_{j} Q_{jj}$, Eq. (11) can be rewritten
as follows $\sigma^{RQE}=\Sigma_{ij}e_{ij}\otimes [U^{QE}{\cal R}
(Q_{ij})U^{QE \dag}]$: Taking partial trace over $R$ leads us to
\begin{eqnarray}
 \sigma^{QE}=
 (ad_{U^{QE}}\circ {\cal R})(\rho^{Q}).
\end{eqnarray}
Tracing out the environment and comparing
the result with Eq. (1) proves the following statement which is the first main point of our study.

\textit{Theorem 1.} If the
initial system-environment state $\rho^{QE}$ is a marginal (reduced) state of a
pure Markov state
then the evolution
of $Q$ is described by
\begin{eqnarray}
  \sigma^{Q}
 ={\cal E}(\rho^{Q}),\quad
 {\cal E}=
 Tr_E\circ ad_{U^{QE}}\circ {\cal R},
\end{eqnarray}
where  ${\cal R}$ is the Petz map given by Eq. (4).
${\cal E}$ is a CPTP map on the support of
initial system state $\rho^{Q}$ for every unitary joint evolutions
$U^{QE}$. $\square$

Partial traces, adjoint actions by unitary operators
(also by isometry operators) and the Petz map are all
CPTP maps, that is
quantum channels. Since concatenation
of quantum channels is again a quantum channel, the evolution map ${\cal
E}$ is a quantum channel on the support of $\rho^{Q}$. We should emphasize
that in the proof of
\textit{Theorem 1} presented in this section no form of
a pure Markov state is used, that is, it is also valid for any
Markov state and with this general form it can be seen from Theorem 1 of Ref.
\cite{Buscemi2014}.

\subsection{Kraus Operators}
In view of Eq. (4) the action ${\cal E}(X)=
 Tr_E\big[ad_{U^{QE}}\circ {\cal R}(X)\big]$ of ${\cal E}$ on any $X\in B({\cal H}_Q)$  can be written as
\begin{eqnarray}
  {\cal E}(X)=
 Tr_E\Big\{ad_{U^{QE}
(\rho^{QE})^{1/2}}\big[\big(ad_{(\rho^{Q})^{-1/2}}X\big)
 \otimes \mathbb{I}_E\big]\Big\}.\nonumber
\end{eqnarray}
Now we consider the
orthonormal basis $\{|\varepsilon_\ell\rangle;
\langle \varepsilon_k|\varepsilon_\ell\rangle=\delta_{k\ell}\}$
 completing the set of  initial
eigenvectors of the  environment to a complete set. Then by writing
$\sum_\ell |\varepsilon_\ell\rangle\langle \varepsilon_\ell|=\mathbb{I}_E$
and evaluating  $Tr_E$
via the same basis we get the Kraus representation   \cite{Kraus}
\begin{eqnarray}
  {\cal E}(X)=\sum_{k,\ell} ad_{E_{k\ell}}(X),
 \end{eqnarray}
where the Kraus operators are obtained as follows
\begin{eqnarray}
 E_{k\ell}=\langle \varepsilon_k|
 U^{QE}
 (\rho^{QE})^{1/2}|\varepsilon_\ell\rangle(\rho^{Q})^{-1/2}.
\end{eqnarray}
It should be emphasized that the first index of $E_{k\ell}$ ranges
over the whole
basis of $E$ and second index takes values in the set of
initial eigenvectors of $E$.

The Kraus operators encapsulate the knowledge
of the initial
system-environment state and of the joint unitary
evolution. It is
well-established general fact that a linear map is CP iff it is a sum
of adjoint actions generated by
a set of Kraus operators. Now, we shall evaluate  the Kraus operators given
by Eq. (15) for pure Markov states.
Making use of
\begin{eqnarray}
(\rho^{QE})^{1/2}&=&
\sum_{j}\sqrt{\kappa_j}|\psi_j^{QE}\rangle\langle \psi_j^{QE}|,\nonumber\\
(\rho^{Q})^{-1/2}&=&\sum_{i,k}(\kappa_i\mu_k)^{-1/2} |q_{ik}\rangle\langle
q_{ik}|
,\nonumber
\end{eqnarray}
and $|\psi_j^{QE}\rangle=\sum_{k}\sqrt{\mu_k}
  |q_{jk}\rangle\otimes|\varepsilon_k\rangle$
 the Kraus operator can be rewritten as
 \begin{eqnarray}
 E_{k\ell}=\sum_{j}\langle \varepsilon_k|
 U^{QE}|\psi_j^{QE}\rangle\langle q_{j\ell}|.
\end{eqnarray}

When Eq. (16) is inserted into
Eq. (14) we obtain
\begin{eqnarray}
  {\cal E}(X)=\sum_{i,j} Tr_Q(X\Pi_{ij})Tr_E\big(
  U^{QE}P_{ij}^{QE}
  U^{QE\;\dag}
  \big),
 \end{eqnarray}
where we have defined
\begin{eqnarray}
 \Pi_{ij}=\sum_{\ell}|q_{i\ell}\rangle\langle q_{j\ell}|,\quad
 P_{ij}^{QE}=|\psi_j^{QE}\rangle\langle\psi_i^{QE}|.
\end{eqnarray}
Noting that $Tr_{QE}P_{ij}^{QE}=\delta_{ij}$ from  Eq. (17)
we get $Tr_Q{\cal E}(X)=
\sum_{i} Tr_Q(X\Pi_{i})$, with $\Pi_{i}=\Pi_{ii}$,
 which implies that ${\cal E}$ is trace-preserving for the
operator defined on the support of $\rho^{Q}$ denoted by $supp(Q)$.
Thus, the trace preserving condition is equivalent
to $\sum_{k,\ell} E_{k\ell}^{\dagger}
E_{k\ell}=\mathbb{I}_{supp(Q)}$. In what follows when
${\cal E}$ is referred to as a channel
this support restriction must be understood.

\subsection{Identification of some channels}
In order to identify some special forms of the channel,
in terms
of the traceless ($Tr_QT(X)=0$) linear map $T$
\begin{eqnarray}
T(X)=\sum_{i\neq j} p_{ij}(X)Tr_E\big(
  U^{QE}P_{ij}^{QE}
  U^{QE\;\dag}
  \big),
\end{eqnarray}
 and the so called Holevo map
 \begin{eqnarray}
  {\cal E}^H(X)=\sum_{i} p_{i}(X)Tr_E\big(
  U^{QE}P_{i}^{QE}
  U^{QE\;\dag}
  \big),
\end{eqnarray}
where $p_{ij}(X)=Tr_Q(X\Pi_{ij}), p_{i}(X)=Tr_Q(X\Pi_{i})$ and
$P_{i}^{QE}=|\psi_i^{QE}\rangle\langle\psi_i^{QE}|$,
  Eq. (17) can be rewritten, more concisely as
\begin{eqnarray}
  {\cal E}(X)={\cal E}^H(X)+
  T(X).
\end{eqnarray}
Eq. (17), or equivalently Eq. (21) provides
 the must general form of the channel implied by the pure
 Markov states. These are given in the bases of initial
 states of $Q$ and $E$.
 In these bases all operators taking part in both
 ${\cal E}^H(X)$ and
 $T(X)$ are independent from $X$. While each operator appearing
 in the former is a state, entirely determined by the joint unitary
 evolution and initial pure $P^{QE}_{i}$ states,
all of operators appearing in the latter are traceless. These forms have
also some remarkable special cases to be mentioned.

In particular;
when
$p_{ij}(X)=Tr_Q(X\Pi_{ij})=0$ for all $i\neq j$, or when $T(X)=0$
the channel has the form ${\cal E}=
{\cal E}^H$.
For this reason we would like to call
the channel of the form (20) the Holevo channel \cite{Holevo1999} which is
known also as the entanglement breaking channel \cite{EBC 2003}.
In that case the output states are convex mixture of $\sigma_i^{ Q}=Tr_E\big(
  U^{QE}P_{i}^{QE}
  U^{QE\;\dag}
  \big)$ such that mixture fractions depends on the input state.

Let us now consider the examples implied by the pure Markov states given by
Eq.  (9). The first
relation
$\rho^{RQE}=|\varphi^{RQ}\rangle\langle
\varphi^{RQ}| \otimes |\psi^{E}\rangle\langle
\psi^{E}|$ of Eq. (9) represents the purification of
initial uncorrelated system-environment
state considered in the majority of the related
literature which provides CPTP evolution for
all joint unitary evolutions. This is a pure Markov state in which the
environment is in the pure state
$|\psi^{E}\rangle$.

On the other hand, in the
product state $|\psi^{RQE}\rangle=|\varphi^{R}\rangle\otimes |\Psi^{QE}\rangle$
of Eq. (9)
we can consider
$|\Psi^{QE}\rangle$ to be a maximally entangled
initial state of $QE$. Obviously the
marginal $\rho^{RE}=|\varphi^{R}\rangle\langle \phi^R|\otimes \mathbb{I}_E/d_E$,  is a
product state and $\rho^{Q}$ and $\rho^{E}$ are maximally mixed. Hence
the associated CPTP map is defined on the whole $B({\cal H}_Q)$.
Since in this case
$R$ is in a pure state, the indices $i$ and $j$ take
only one value and therefore can be suppressed to obtain
\begin{equation}
{\cal E}(X)= Tr_Q(X)Tr_E\big(
  U^{QE}P^{QE}
  U^{QE\;\dag}
  \big),\nonumber
\end{equation}
where $P^{QE}=|\psi^{QE}\rangle\langle\psi^{QE}|$.
  Thus ${\cal E}$ is a completely
  depolarizing channel mapping all states to a fixed state
$\sigma^{Q}=Tr_E\big(
U^{QE}P^{QE}
U^{QE\;\dag}
\big)$. Note that this is a special case of the Holevo channel.

The rest of this study is devoted to characterization
of the initial system-environment correlations.
It turns out that the entropy $S(E)$ of the environment play a vital
role in this context such that
whenever it is nonzero initial states of $QE$ are correlated.

\section{Initial system-environment correlations}
Our main goal in this section is to specify
both quantitatively and qualitatively  quantum correlations and
classical correlations of the joint system QE when it is in a marginal state of a pure Markov
state.
 The emphasize will be put on the entanglement of formation,
discord and classical correlations and we firstly recall
their definitions for a generic
bipartite state $\rho^{AB}$. Then, relationships between
these information theoretical
quantities will be established for general tripartite
pure states and finally the
relationships for the pure Markov states will be deduced from them.

\subsection{Entanglement of formation (EOF)}
When $\rho^{AB}$ is a pure state we have
$S(AB)=0$ and $S(A)=S(B)$, hence the entropy of
a marginal state is a natural quantitative
measure of the entanglement of a bipartite pure state. A
given bipartite pure state is
entangled iff the entropy of its marginal states are
different from zero and it is
maximally entangled iff the entropy of its
marginal states are maximum \cite{NC,Holevo}. For a given mixed state
$\rho^{AB}$ there is not so easy way of even deciding the
existence of entanglement. Perhaps
the most efficient way is to define, again by means of the entropy of
a marginal state, say $\rho^{A}$, the
EOF of $\rho^{AB}$ by \cite{Bennett,Wootters}
\begin{eqnarray}
 E_f(AB)=\inf_{\{p_i,P_i^{AB}\}}\sum_ip_iS(Tr_B P^{AB}_i),
\end{eqnarray}
where positive numbers $p_i$ denotes the probabilities $\sum_ip_i=1$ and
$P^{AB}_i=|\psi^{AB}_i\rangle\langle \psi^{AB}_i|$ are rank-1
projectors (pure states) such
that  $\langle \psi^{AB}_i|\psi^{AB}_i\rangle=1$ for all $i$, but
they do not need to be
orthogonal. The \textit{infimum} in Eq. (22) ranges over all possible pure-state
decompositions of $\rho^{AB}=\sum_ip_iP^{AB}_i$.

\subsection{An equivalent form of the EOF}
For our purpose in this study we should
convert the conventional definition of EOF to another equivalent form such that it will be
possible to deal with EOF and other measures of correlations on equal footing. For this
purpose we shall firstly replace the range of infimum of Eq. (22) with an equivalent set and
secondly the average entropy of Eq. (22) will be replaced with a more suitable conditional
entropy. To accomplish the first replacement, we note that any pure-state decomposition of a
given density matrix can be obtained as the non-selective rank-1 POVM measurements locally
carried out on its  purifying system. A POVM ${\cal M}$ is a collection of a complete set of
the POVM elements $M_i$ each of which is positive operator and is associated with a single
measurement.

Let $P^{RAB}=|\Phi^{RAB}\rangle\langle \Psi^{RAB}|$ be a purification of a given $\rho^{AB}$
and let $\rho^{AB}=\sum_ip_iP^{AB}_i$ be a pure state decomposition of it.
We shall denote the
rank-1 POVM that produces this decomposition via its
nonselective action on $R$ by the set
\begin{equation}
{\cal M}=\{0\leq M_i\leq \mathbb{I}_R;\sum_iM_i=\mathbb{I}_R\}.
\nonumber
\end{equation}
Rank-1 POVM means that
$M_i=|\alpha_i\rangle\langle\alpha_i|$ for all $i$ such that
$|\alpha_i\rangle$'s  need to be
neither normalized nor orthogonal. The explicit form of the correspondence between the pure state
decomposition and complete local execution of the POVM on the purifying reference system $R$
can be written as
\begin{eqnarray}
 \rho^{AB}&=&\sum_ip_iP^{AB}_i\nonumber\\
 &=&\sum_iTr_R\big[(\sqrt{M_i}\otimes \mathbb{I}_{AB})
 P^{RAB}(\sqrt{M_i}\otimes \mathbb{I}_{AB})\big],\nonumber
\end{eqnarray}
where $\rho^{R}=Tr_{AB}P^{RAB}$ and $p_i=Tr_R(M_i\rho^{R})$. Rank-1 condition for all the POVM
elements $M_i$'s is sufficient (as well as necessary) for the purity of all conditional states
$P^{AB}_i$ (see Appendix A where this statement and its converse are proved).

For the second replacement in the definition of EOF, we observe
that in terms of an
orthonormal basis $\{|i\rangle\}$ of an auxiliary
Hilbert space ${\cal H}_X$ to any pure-state
decompositions (in fact, for any convex mixture)
of $\rho^{AB}$ is associated a
classical-quantum state $
 \rho^{XAB}=\sum_ip_i|i\rangle\langle i|\otimes P^{AB}_i.$
In such a case, since $\rho^{A}_i=Tr_BP^{AB}_i$ the average entropy
$\sum_ip_iS(\rho^{A}_i)$
appearing in the definition (22) is nothing more than the
conditional entropy $S(A|X)$ of the
state $\rho^{XA}=\sum_ip_i|i\rangle\langle i|\otimes \rho^{A}_i$. Thus
we can rewrite EOF
given by (22) as follows
\begin{eqnarray}
E_f(AB)=\inf_{{\cal M}}\big\{S(A|X);
 \rho^{XA}=\sum_ip_i|i\rangle\langle i|\otimes \rho^{A}_i \big\},
\end{eqnarray}
where $p_i\rho^{A}_i=Tr_{RB}[(M_i\otimes \mathbb{I}_{AB})P^{RAB}].$
This is our promise for the new equivalent definition of EOF and it should be emphasized that
here the \textit{infimum} must be taken over all rank-1 POVMs acting on $R$.

\subsection{Discord and classical correlations}
The classical correlation
$C(R\rightarrow B)$ between $R$
and $B$ in the bipartite state $\rho^{RB}=Tr_AP^{RAB}$
can be specified by means of the so called Holevo
quantity
\begin{eqnarray}
\chi(\{p_i,\rho^{B}_i\})=S(\sum_ip_i\rho^{B}_i)-\sum_ip_iS(\rho^{B}_i),\nonumber
\end{eqnarray}
of the state $\rho^{B}=Tr_{RA}P^{RAB}=\sum_ip_i\rho^{B}_i$.
The Holevo
quantity is a fundamental upper bound for the
accessible information between the sender and a receiver
communicating classical messages by quantum means
(Theorem 5.9 of the Ref. \cite{Holevo}). Here $\rho^{B}_i$ is
the carrier of
the classical message
$i$ send with probability $p_i$  and the receiver tries to read the
message by POVM measurements. A fundamental relation between the
classical correlations and EOF can be stated as follows.

\textit{Lemma 3.} For any  purification $\rho^{RAB}$
of a given bipartite state
$\rho^{AB}$, the EOF $E_f(AB)$ for $\rho^{AB}$ and
the classical correlations $C(R\rightarrow
B)$ between $R$ and $B$ always obey the following equality
\begin{eqnarray}
C(R\rightarrow B)+E_f(AB)=S(B).
\end{eqnarray}
This relation is well known in the literature as the
Koashi-Winter \textit{monogamy} relation \cite{Koashi-Winter2004} for which
an
alternative and instructive proof can be given as follows.

\textit{Proof.} In terms of the associated classical-quantum state
$\rho^{XB}=\sum_ip_i|i\rangle\langle i|\otimes \rho^{B}_i,
\chi(\{p_i,\rho^{B}_i\})$
can be written as
\begin{eqnarray}
\chi(\{p_i,\rho^{B}_i\})=S(X;B)=S(B)-S(B|X).\nonumber
\end{eqnarray}
Now, the classical correlations $C(R\rightarrow B)$  are defined
by maximizing the Holevo quantity  over all
rank-1 POVMs executed on $R$;
\begin{eqnarray}
C(R\rightarrow B)=S(B)-\inf_{{\cal M}}S(B|X).
\end{eqnarray}
Since $E_f(AB)=E_f(BA)$ the second term at the right hand side is the
EOF given by Eq. (23). This, proves Eq. (24). $\square$

For later convenience in adopting these relations for the pure Markov states, we simply change
the  label of systems  $A$ and $B$, respectively by $Q$ and $E$ and
consider an arbitrary pure state $\rho^{RQE}$ which need not
be a Markov state yet.
For this case we rewrite Eq. (24) as
\begin{eqnarray}
C(R\rightarrow E)+E_f(QE)=S(E).
\end{eqnarray}
Making use of Eq. (2) and  the relations implied by the bipartite splits
of $\rho^{RQE}$ we have
\begin{eqnarray}
 S(R)&=&\frac{1}{2}[S(R;Q)+S(R;E)],\nonumber\\
 S(Q)&=&\frac{1}{2}[S(R;Q)+S(Q;E)],\\
 S(E)&=&\frac{1}{2}[S(R;E)+S(Q;E)].\nonumber
\end{eqnarray}
These simply state that entropy of any individual part of any tripartite
pure state is the
arithmetic mean of mutual information of that part with remaining other two partners.
Recalling the
definition of discord \cite{discord}
\begin{eqnarray}
D(Q\rightarrow E)=S(Q;E)-C(Q\rightarrow E)
\end{eqnarray}
between $Q$ and $E$ we can write two important corollaries of \textit{Lemma 3}.

\textit{Corollary 2.} For any  purification $\rho^{RQE}$ of a given bipartite state
$\rho^{QE}$, the following equalities hold:
\begin{eqnarray}
C(Q\rightarrow E)+E_f(RE)&=&S(E),\\
S(R;E)-E_f(RE)+D(Q\rightarrow E)&=&S(E).
\end{eqnarray}
\textit{Proof.} The first relation is obtained from Eq. (26)
by the
interchange $R\leftrightarrow Q$ and
the second relation is the difference of third relation
of Eq. (27) and Eq. (29). $\square$

\subsection{EOF, discord and classical correlations implied by the pure Markov states}
Having specified the quantum mechanical as well as the classical correlations
of any pure tripartite state in terms of information theoretical quantities
we are ready to apply these results to any pure Markov state $\rho^{RQE}$.
Since $\rho^{RE}$ is a product state in this case the mutual information,
the classical correlations and the EOF between $R$ and $E$ are zero:
\begin{eqnarray}
S(R;E)=0=E_f(RE)=C(R\rightarrow E).\nonumber
\end{eqnarray}
Thus Eqs. given by (26), (29) and (30) reduce to the following forms
\begin{eqnarray}
E_f(QE)=C(Q\rightarrow E)=D(Q\rightarrow E)=S(E).\nonumber
\end{eqnarray}
That is, numerical values of considered correlations between $Q$ and $E$
are equal to each other and their value
is simply
given by the entropy of the environment.
Moreover Eqs. (27) take the forms
\begin{eqnarray}
 S(R)=\frac{1}{2}S(R;Q),\quad S(E)=\frac{1}{2}S(Q;E),\nonumber
\end{eqnarray}
and $S(Q)=[S(R;Q)+S(Q;E)]/2$ for any pure Markov state.
These observations can be summarized
by the following theorem which emphasizes another general
trait of the pure Markov states.

\textit{Theorem 2.} For any  pure Markov state $\rho^{RQE}$, conditional on $Q$;
 the EOF, discord and classical correlations between the
 system $Q$ and its environment $E$
are all numerically equal to each other and to
the entropy of the environment $E$ which is one-half of the
mutual information of $Q$ and $E$.$\square$

Eqs. (26) and (29) represent a kind of conservation law for the correlations
of an environment
with the related OQS and purifying system. For a given
$S(E)$ both
$C(R\rightarrow E)$ and $E_f(QE)$ (or $C(Q\rightarrow E)$ and $E_f(RQ)$)
can change (without exceeding the value $S(E)$)
but their sum is always fixed by $S(E)$. In the case of a pure Markov
state both $C(Q\rightarrow E)$ and $E_f(QE)$ are equal to $S(E)$ which means
that as long as $S(E)\neq 0,\;Q$ and $E$ are certainly correlated and
even in such a case the reduced dynamics of $Q$
can be described by a CPTP map.
\section{CONCLUDING REMARKS}
Finally we would like to present our remarks concerning two seemingly
different topics; \textbf{(i)} perfect quantum error correction
(QEC) in quantum information theory
\cite{BS and Nielsen96-2002} and  \textbf{(ii)}
quantum non-Markovian (or Markovian) dynamics in the
theory of OQSs \cite{Breuer-Petruccione2002,Rivas Huelga 2014}.

\textbf{(i)} Main point of the perfect QEC procedure is to
faithfully restore the output state of a
system $Q$ to its
input state
by means of local
 transformations executed at the receiver side after $Q$ has been sent
 through a noisy transmission channel. For the transfer of
 quantum information
 the input is supposed to be a part of an entangled pure state
 with a reference system $R$
 and the noisy channel is modelled
 to arise, as usually, from a joint unitary evolution of $Q$ and
 its environment $E$ which is
 supposed to be in a pure state \cite{BS96}. During the
transmission process, $Q$ evolves as an open system and
 $R$ remains
 intact throughout the process. Thus
 tripartite framework and uncorrelated
 initial $QE$ state are essential in the present
 perfect QEC scheme. Moreover, the initial tripartite
 state is a special pure Markov state corresponding to
 our first relation of Eq. (9). What is more,
 being another pure Markov state for the tripartite output
 suffices for the accomplishment of perfect QEC
 \cite{BS and Nielsen96-2002}.
 Hence perfect QEC procedure can be formulated as preserving
 the pure Markov structure
 of the tripartite input state and we do naturally expect that the results
 of this study can provide a broader perspective
 on the perfect QEC and on the approximate QEC procedures.

 \textbf{(ii)} Quantum
non-Markovianity (and Markovianity) is another central
topic in the theory of OQSs.
This can be defined via
 divisibility property of evolution
 maps \cite{Rivas Huelga 2014}. A quantum system $Q$
 subject to a time evolution given by
some family of trace-preserving linear maps $\{{\cal
E}_{\tau^{\prime},\tau};\tau^{\prime}\geq\tau\geq \tau_0\}$ is Markovian
(or divisible) iff,
for every $\tau^{\prime}$ and $\tau$ , ${\cal E}_{\tau^{\prime},\tau}$ is
a CPTP
map and fulfils the composition law
\begin{eqnarray}
{\cal E}_{\tau^{\prime\prime},\tau}=
 {\cal E}_{\tau^{\prime\prime},\tau^\prime}\circ {\cal E}_{\tau^{\prime},\tau}
 ,\quad \tau^{\prime\prime}\geq\tau^\prime \geq \tau.
\end{eqnarray}
To establish a concrete connection with our work we restore time labelling
 of initial and final states respectively by $\tau$ and $\tau^\prime$ as
 mentioned in Section II.
As we have shown, even in the presence of initial
correlations the existence of CPTP evolution
${\cal E}_{\tau^{\prime},\tau}$ is guaranteed as long as
the tripartite initial state
is a pure Markov state.
Although the tripartite output at time $\tau^\prime$ is a pure state
$\sigma^{RQE}_{\tau^{\prime}}= |\phi_{\tau^{\prime}}^{RQE}\rangle\langle
\phi_{\tau^{\prime}}^{RQE}|$ where
$|\phi_{\tau^{\prime}}^{RQE}\rangle=(\mathbb{I}_R\otimes
U^{QE}_{\tau^{\prime},\tau})|\psi_\tau^{RQE}\rangle$, it need not be a
pure Markov state. Hence there may not be a CP map
${\cal E}_{\tau^{\prime\prime},\tau^\prime}$ after the time $\tau^\prime$.
According to the \textit{Corollary 1}, a sufficient conditions
for this to be case is that the
reduced output
\begin{eqnarray}
 \sigma^{RE}_{\tau^{\prime}}=
 Tr_Q\big [id_R\otimes (ad_{U^{QE}_{\tau^{\prime},\tau}}\circ
 {\cal R})\big]\rho_\tau^{RQ}
\end{eqnarray}
being a product state. Whenever this occurs the existence of CPTP map
${\cal E}_{\tau^{\prime\prime},\tau^\prime}$ will be guaranteed.
Thus, we can say that
an evolution which preserves the pure Markov structure, that is,
transforming a pure Markov state to another one for
all $\tau^\prime$ is a
Markovian evolution for the observed system.

To sum up, the problems such as the reduced quantum dynamics in the presence
of initial system-environment correlations, the Markovian and non-Markovian
dynamical evolutions
and the perfect QEC and approximate QEC procedures are very intimately
connected. Pure Markov states, or in a broader context Markov states play a
central role in these seemingly different topics.
\begin{acknowledgments}
We are grateful to  anonymous referee whose valuable suggestions and comments improved the paper.
This work was supported in part by the Scientific and Technological Research Council of Turkey
(TÜBİTAK).
\end{acknowledgments}
\begin{appendix}
\section{}
Let $|\Upsilon^{AR}\rangle$ be a purification of a
given state $\rho^{A}$ such that $\rho^{A}=Tr_RP^{AR}$ with
$P^{AR}=|\Upsilon^{AR}\rangle\langle \Upsilon^{AR}|$ and let $
{\cal M}=\{0\leq M_i\leq \mathbb{I}_R;\sum_iM_i=\mathbb{I}_R\}$,
be
a POVM on the purifying system $R$. The conditional state $\rho_i^{A}$,
known also as
the post-measurement state,
resulting from a single local measurement is given by
\begin{eqnarray}
p_i\rho_i^{A}
 =Tr_R\big[(\mathbb{I}_{A}\otimes\sqrt{M_i})
 P^{AR}(\mathbb{I}_{A}\otimes \sqrt{M_i})\big],
\end{eqnarray}
where $ p_i=Tr_R(M_i\rho^{R})$ is nonzero and $\rho^{R}=Tr_{A}P^{AR}$
.
\textit{Lemma A.} The conditional
state $\rho_i^{A}$ defined by (A1) is a pure state iff
the POVM element $M_i$ is of rank-1 \cite{Note3}.

\textit{Proof.} Let $\{|\psi_m\rangle\}$ be the set of the
orthonormal eigenvectors of  $\rho^{A}$
corresponding to nonzero eigenvalues $\{\lambda_m\}$ and
let $\{|m\rangle\}$ be an orthonormal
basis of $R$. By writing
\begin{eqnarray}
P^{AR}=\sum_{mn}\sqrt{\lambda_m\lambda_n} |\psi_m\rangle\langle \psi_n|\otimes
|m\rangle\langle n|,\nonumber
\end{eqnarray}
we have, from (A1)
\begin{eqnarray}
p_i\rho_i^{A}
 =\sum_{mn}\sqrt{\lambda_m\lambda_n}|\psi_m\rangle\langle \psi_n|Tr_R\big
 ( M_i|m\rangle\langle n|\big).
\end{eqnarray}
When $M_i$ is of rank-1 we can write $M_i=|\alpha_i\rangle\langle \alpha_i|$ where nonzero
$|\alpha_i\rangle$ need not be normalized. Then by defining $c_{im}=\langle
\alpha_i|m\rangle$  from (A2)
we have
$p_i\rho_i^{A}=
\sum_{mn}c_{im}c^{\ast}_{in}\sqrt{\lambda_m\lambda_n}|\psi_m\rangle\langle \psi_n|$,
where $^\ast$ denotes the complex conjugation, and in terms of
$|\Phi_i\rangle=\sum_{m}c_{im}\sqrt{\lambda_m} |\psi_m\rangle$ we
obtain  $p_i\rho_i^{A}=|\Phi_i\rangle\langle
\Phi_i|$. Since
$\rho^{R}=\sum_{m} \lambda_m|m\rangle\langle m|$ and
\begin{eqnarray}
p_i=Tr_R(M_i\rho^{R})=\sum_{m}\lambda_m|c_{im}|^{2}=\langle
\Phi_i|\Phi_i\rangle,
\nonumber
\end{eqnarray}
in terms of the normalized state $|\varphi_i\rangle
=|\Phi_i\rangle/\sqrt{p_i}$ we have $\rho_i^{A}=|\varphi_i\rangle\langle\varphi_i|$.
Conversely, suppose that $\rho_i^{A}$ is a pure state such that
 $\rho_i^{A}=|\beta_i\rangle\langle \beta_i|$ with $\langle
\beta_i|\beta_i\rangle=1$. Then from (A2) we obtain
\begin{eqnarray}
p_i|\beta_i\rangle\langle \beta_i|
 =\sum_{mn}\sqrt{\lambda_m\lambda_n}|\psi_m\rangle\langle \psi_n|
 (M_i)_{nm}.
\end{eqnarray}
Here $(M_i)_{nm}=\langle n|M_i|m\rangle$
denotes the matrix element of $M_i$ in the orthonormal basis
$\{|m\rangle\}$ of $R$. In terms
of $
b_{ik}=\sqrt{p_i/\lambda_k}\langle \psi_k|\beta_i\rangle
$
from
(A3) we obtain $(M_i)_{k\ell}
=b^{\ast}_{i\ell}b_{ik}$ which implies $M_i=|b_i\rangle\langle b_i|$, with $\langle
b_i|m\rangle=b_{im}$. $\square$
\end{appendix}

\end{document}